# Topography, astronomy and dynastic history in the alignments of the pyramid fields of the Old Kingdom.

Giulio Magli
Faculty of Civil Architecture - Politecnico di Milano
Piazza Leonardo da Vinci 32, 20133 Milan, Italy

*It is known since the 19 century that in the layout of the pyramid field of the pharaohs of the 4$^{th}$ Egyptian dynasty at Giza, a "main axis" exists. Indeed, the south-east corners of these monuments align towards the site of the temple of Heliopolis, which was plainly visible in ancient times. It was later discovered that a similar situation occurs in the main pyramid field of the subsequent dynasty at Abu Sir. Here, the north-west corners of three chronologically successive pyramids again voluntarily align towards Heliopolis. However, the temple was in this case not visible, due to the rock outcrop- today occupied by the Cairo citadel - which blocks the view. In the present paper, a inter-disciplinary approach based on historical, topographical and archaeoastronomical analysis is developed in an attempt at understanding such peculiar features, which governed from the very beginning the planning of these wonderful monuments. A general pattern actually arises, which appears to have inspired the choice of the sites and the disposition on the ground of almost all the funerary complexes of the kings during the Old Kingdom. In particular, this pattern helps to explain the choices in the location of the funerary complexes of Niuserre in Abusir and of Unas in Saqqara.*

# 1. Introduction

An interesting feature exists in the layouts of the pyramids of Giza and Abu Sir: the presence of a "main axis" directed to the area where the ancient temple of the sun of Heliopolis once stood, on the opposite bank of the Nile. These axes are connected with a process of "solarisation" of the pharaoh which probably started with Khufu, the builder of the Great Pyramid. At Giza, the axis (already discovered in the 18[th] century) runs across the south-east corners of the main pyramids (Goedicke 2001, Lehner 1985a,b, Magli 2009a,b). In Abu Sir, a straight line connects the north-west corners of the pyramids of three successive kings (Verner 2002). However, the view to Heliopolis is *blocked* here by the rock outcrop of the Cairo citadel (Jeffreys 1998). This "alignment" is, therefore, quite mysterious: its intentionality and relationship with the cult of the sun are certain, but its aim seems to be failed. As a consequence, the very choice of the site of Abusir by the kings of the 5[th] dynasty remains unexplained (see discussions in Krejci 2001, Goedicke 2001).

The research presented here originally started as an attempt at solving this enigma; however, it turned out that what was necessary was a careful topographical and archaeoastronomical analysis of the *whole* set of monuments built between the 4[th] and the 6[th] dynasty, accurately taking into account what was known historically and archaeologically about them. In the course of such an analysis, a general pattern was seen to arise, which inspired the location and planning of almost all monuments and *in particular* led to the choice of the Abu Sir plateau. This pattern - which I will call, for a reason which will be clear soon, *symbolic invisibility* - is based on a set of radial lines ("diagonals") which connect in a ideal way Heliopolis with the sacred landscape built on the opposite bank of the Nile. Its origins appears to be connected with the ancient Egyptians' religion, funerary cult and dynastic lineage, as well as with their astronomical knowledge. Further, as we shall see, the idea of connecting a new funerary monument with the existing ones by means of "topographical alignments" remained alive also in the later times of the Old Kingdom, during the end of the 5[th] dynasty and the run of the 6[th]. It lost, however, the "solar" character of connection towards Heliopolis.

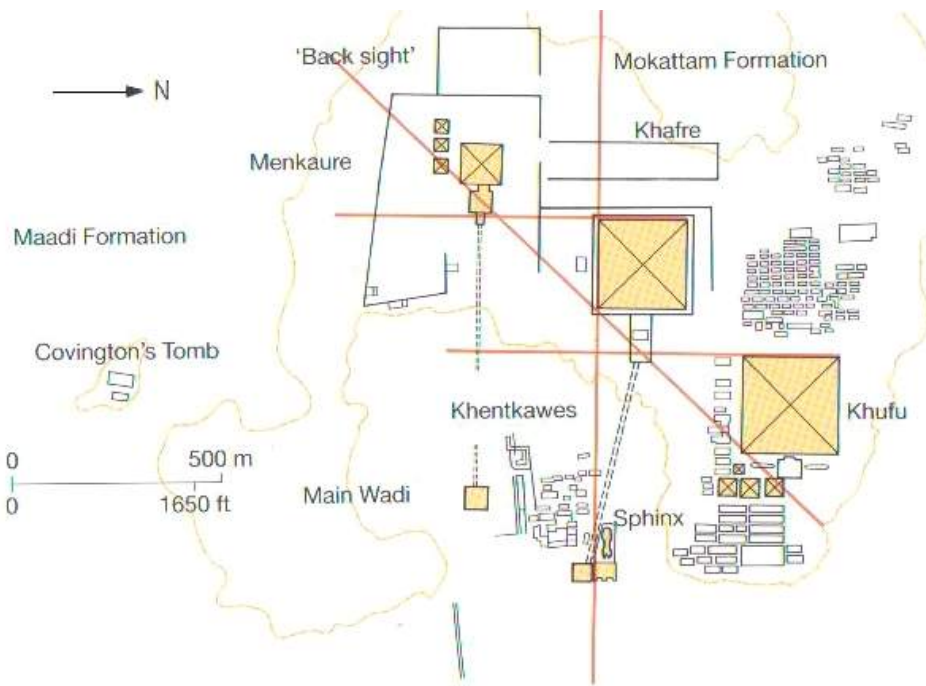

**Fig. 1 Map of the main Giza pyramids with the Giza axis highlighted (from Lehner 1999)**

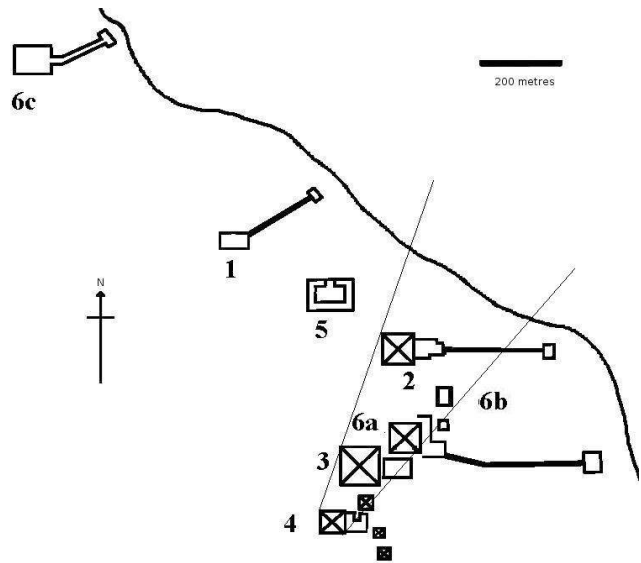

**Fig. 2** Map of the Abu Gorab - Abu Sir area with the "Abu Sir diagonals" highlighted (numbering of the monuments in chronological order). 1- Userkaf Sun temple, 2/3/4 Pyramids of Sahure, Neferirkare, Neferefre 5-unfinished pyramid of Shepseskare 6a/6b/6c Niuserre Pyramid, Mastaba of Ptahshepses, Niuserre Sun temple.

## 2. The geography of the Memphite area and the horizon formula

The geographical area of interest in the present paper can be visualized as a triangle (Fig. 3). The western leg of the triangle follows the rock ridge of the desert on the west bank of the Nile from Abu Roash (A) to Abu Gorab (B) crossing Giza (G) and Zawiet el Arian (Z), while the other two legs are defined by the position (H) of ancient Heliopolis, located on the east bank of the Nile (to visualize this area I have chosen as A the center of Djedefre Pyramid, as B the center of the Userkaf Sun Temple and as H the position of the Middle Kingdom obelisk in the Heliopolis area). The leg H-A is about 25 Kms long and crosses the flatland of the Nile valley; the leg H-B is about 27.5 Kms long; it also crosses the Nile flatland but it is "tangent" to the rock outcrop located at the north-west extreme of the Moqattam formation. This outcrop is today occupied by the Cairo citadel (K). Important in what follows will be also Abu Sir (S) which is a plateau located on the ridge of the desert some hundreds of meters south of Abu Gorab, the pyramid field of Saqqara (Q) further south, and the site of the Old Kingdom capital Memphis (M).

We shall concentrate here on the issue of mutual visibility/invisibility between Heliopolis and the above-mentioned sites. Since it is easy to find non-scholarly publications where ancient monuments are allegedly supposed to be connected by "invisible lines", it is perhaps worth clarifying that the present work has nothing to do with most such speculations. Usually indeed they have little sense either because the supposed lines are drawn on a plane chart (while the earth happens to be round) or because they connect sites which do not have any historical connections whatsoever (or because of both reasons together). In the present paper, monuments built in a relatively short period of time by dynastically related rulers are discussed, and the specific issue of inter-visibility is carefully addressed every time a interconnecting line is proposed. To this aim, we first notice that, as mentioned, the Moqattam formation blocks the view between H and any zone further south than Abu Gorab, the first of the invisible zones being the Abu Sir area. The leg HB is therefore the last possible inter-visibility line from H to the west bank looking south. The triangle HAB includes only flat land and therefore there exists a theoretical inter-visibility between all its points, in particular between H and places on the A-B leg. Due to the relevant distances which come into play however, we have to take into account earth's curvature. To this aim, we briefly recall the so-called *horizon formula* (actually a straightforward consequence of Pythagoras' theorem, see Fig. 4).

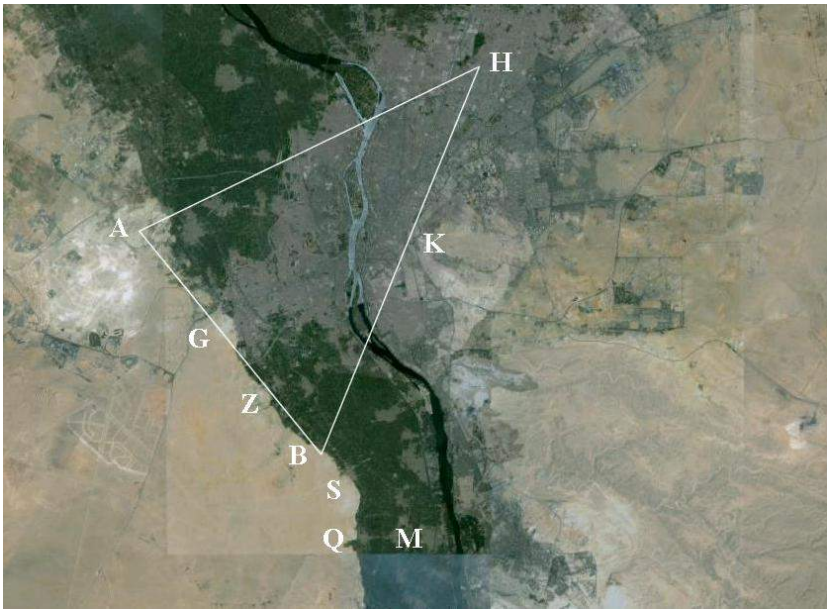

**Fig. 3 The geographical area of interest. H=Heliopolis, K= Cairo Citadel, A=Abu Roash, G=Giza, Z=Zawiet el Arian, B=Abu Gorab, S=Abu Sir, Q=Saqqara main field, M=Memphis**

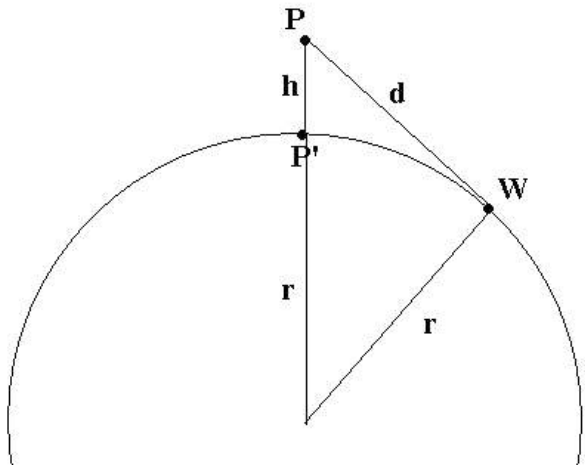

**Fig. 4 The horizon formula**

Given a point P located at vertical height h with respect to the earth surface P', the farthermost point W that can be seen from P is the point where passes the tangent to the earth circumference from P. Clearly then the distance d=PW equals $\sqrt{2rh+h^2}$ where r is the earth radius. If the heights are very small with respect to r we can put $d \sim \sqrt{2rh}$. Since 2r~13000 Km, we arrive at the so-called "horizon formula": the value of d in Kms approximately equals the square root of 13h if h is expressed in meters. It follows, for instance, that a person 2 meters tall has a visible horizon slightly greater than 5 Kms. However, if the object sighted has a non negligible height, then the two sum up. As a consequence, a sun-reflecting signal located in Heliopolis at - say - 20 meters above ground would have been easily visible from the west ridge, located at an average of some 20 meters above the plain. Such a signal might have been located e.g. on the "Petrie's mound", a symbolic "primeval mound" which was probably there since the first dynasties (for a complete discussion on Heliopolis see Quirke 2001 and references therein) or, more simply, on an observation post located at the top of a provisional wooden structure. Once the building sites of the pyramids reached a sufficient height, the upper courses became inter-visible with Heliopolis. The fascinating view of the Giza pyramids from Heliopolis was still enjoyable at the end of the 19-century, as confirmed by some paintings of that times; attempts at a direct observation are doomed to failure today, due to modern buildings and pollution which strongly limits visibility.

## 3. Geographical and chronological order of the monuments

In what follows we shall be interested in almost *all* the pyramidal complexes constructed during the Old Kingdom. The geographical location of these monuments and their chronological order will both be essential (Fig. 5). Before going further however, it is worth recalling that the chronology of the Old Kingdom is subject to debates about the length of each single reign and the succession of a few of the kings. As a reliable working framework I follow here strictly the accession dates as given by Baines and Malek (1984) and adopted, for instance, by Lehner (1999).

The "Age of the Pyramids" starts with the Step Pyramid in Saqqara, built by the first $3^{th}$ dynasty pharaoh Djoser (2630-2611). After him, other pyramids will be built by kings of the same dynasty: the first is Sekhemkhet's in Saqqara After, the so-called Layer Pyramid (Z1) was built at Zawiet el Arian and another very far south, in Meidum. The attribution of these two is uncertain; the first might have been built by a king called Khaba, the second by Huni or by Snefru in the first years of his reign. Also uncertain is the attribution of an unfinished, partly mud-brick pyramid which is located in Abu Roash, the so called Lepsius 1 pyramid (A1); anyhow the substructure excavated in the rock clearly points to the end of the $3^{th}$ /beginning of the $4^{th}$ dynasty (Swelim 1983). The $4^{th}$ dynasty begins with Snefru (2575 - 2551), the builder of the two magnificent pyramids located in Dashour (South Saqqara). With the son of Snefru, Khufu, we enter into the period which is of direct interest for us.

1) Khufu (2551 - 2528) builds the Great Pyramid (G1) in Giza.
2) Djedefre (2528-2520) builds his pyramid (A2) in Abu Roash. Djedefre adds the solar "suffix" -re to his name, formalizing a process of "solarisation" of the king which was certainly initiated by his father Khufu.
3) Khafre (2520 - 2494) builds or finishes building the second pyramid of Giza (G2) (there are recent hints pointing to the fact that the original Khufu project enclosed also this pyramid and Khafre finished it (Shaltout et al. 2007, Magli 2003, 2009), but this is not of special relevance here).
4) Someone, perhaps an undocumented short-reign king, started the building of a huge project, comparable to that of G1, in Zawiet el Arian (Z1).
5) Menkaure ( 2490 - 2472) builds the third pyramid of Giza (G3).
6) Shepsekaf (2472-2467) appears to break the tradition. Indeed he does not build a pyramid, but a giant Mastaba at Saqqara South (called today Mastaba El Faraun); the existence of a break in the funerary and religious customs is reflected also in the pharaoh's title, which does not bring the "solar" suffix (for a discussion of different viewpoints see Verner 2002) .

7) Userkaf (2465 - 2458) returns to the tradition. He builds a pyramid in Saqqara, significantly located as close as possible to the wall of the first pyramid ever constructed, Djoser's. Userkaf also builds the first sun temple in the zone of Abu Gorab (B1). This temple was composed, like the "standard" pyramid's two-temples/causeway complex, by a building set lower down and a monumental access ramp sloping upwards, giving access to a monumental area. Here, a huge stone basement was sustaining a high (non-monolithic) obelisk. Such a monument was certainly devoted to the Sun God and its plan was probably conceived as a copy of the temple in Heliopolis. This place was indeed an extremely important religious center; it was called *Iunu*, pillar, and was a sort of "umbilicus mundi" of the country since the first dynasties.
8) Sahure (2458 - 2446). With Sahure, whose name means "Close to Re", we return to "solarised" kings. The pyramid of Sahure is located in Abu Sir (S1).
9) Neferirkare (2446 - 2426) builds his pyramid in Abu Sir (S2)
10) Neferefre builds his (unfinished) pyramid in Abu Sir (S3)
11) Shepseskare probably builds his pyramid in Abu Sir (S4) as well (the attribution of this unfinished monument is not certain). It is not completely clear if really this king was the successor of Neferefre or rather vice-versa, although, as we shall see, the results of the present paper strongly suggest the first option. In any case both reigns were very short, lasting as a whole 10 years or less (2426-2416).
12) Niuserre (2416-2388) builds his pyramid in Abu Sir (S5) *and* a sun temple (B2) in Abu Gorab. Niuserre was the last king to build his funerary complex in Abu Sir. The last kings of the 5$^{th}$ dynasty will indeed return to Saqqara, as well as all the subsequent pharaohs (Section 8**).

## 4. Visibility, invisibility, and symbolic invisibility

The above list may, at a first glance, lead to conclude that each king decided in a somewhat random way to put his pyramid in one of the available sites on the west bank of the Nile, provided just that it was not too far from the capital Memphis. Actually, the placement of pyramidal complexes did have to take into account a series of practical factors (Barta 2005) such as presence of nearby stone outcrops - to be transformed into quarries - and accessibility of materials. Perhaps also the presence of the - still active - building site of the pyramid of the preceding pharaoh may have influenced the choice. *However*, there is no possible doubt on the fact that, at least in many cases, the pyramids were *not* constructed were reasonableness would have wanted. A clamorous example is that of the Menkaure pyramid (G3), which has been voluntarily built very far in the desert; similarly (and indeed for the very same reason, as we shall see) this holds for the Neferefre pyramid (S3). More generally, it must be remembered that we are speaking here about nothing less than some of the most magnificent, complicated and huge buildings of the whole human history. We do know that such monuments were designed with maniacal accuracy (especially during the 4$^{th}$ dynasty) and that their construction deserved the efforts of thousands of specialized, well feed, skilled workers (Lehner 1999, Hawass 2006). The pyramids were the symbol of the power of the pharaoh over the death and, in reflection, of the continuity of life of the whole country. It is therefore natural to think that their placement was carefully chosen in order to satisfy also to a set of religious, symbolic criteria.

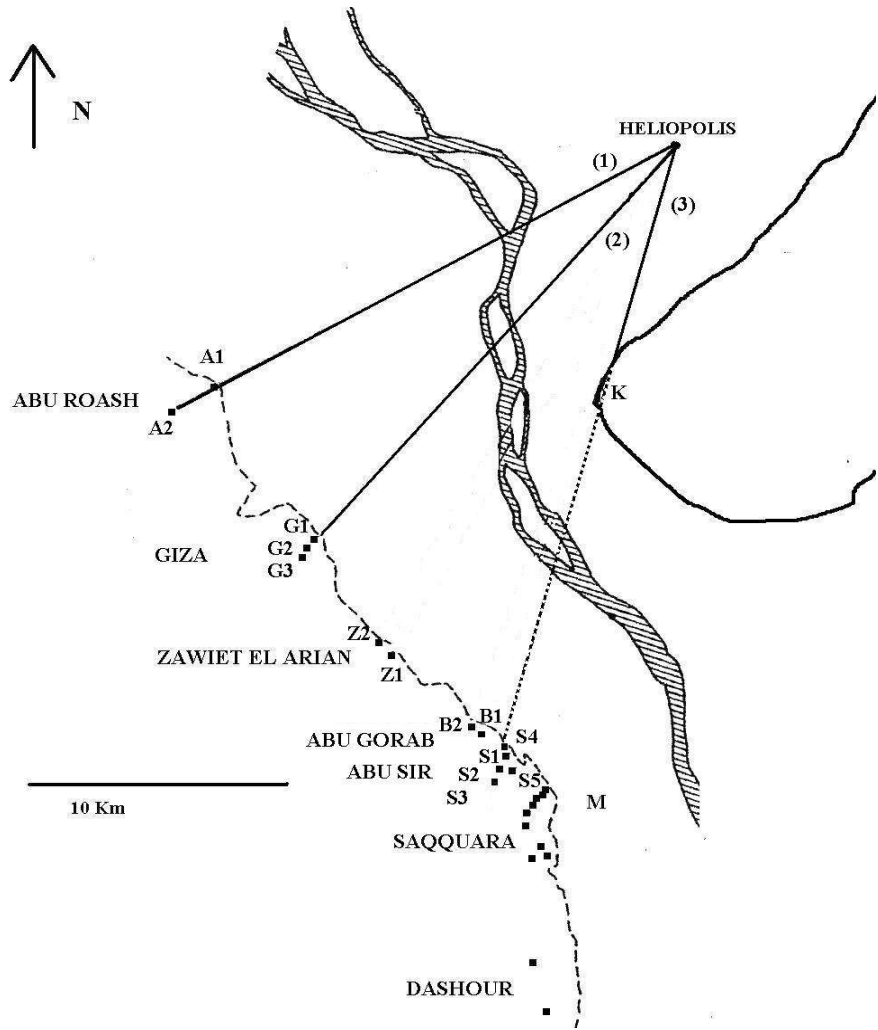

**Fig. 5** The pyramids, the Sun Temples and the Heliopolitan "diagonals" discussed in the paper: (1) Abu Roash diagonal, (2) Giza diagonal, (3) Abusir diagonal. The dotted straight line denotes invisibility. The stepped line denotes the maximal Nile flood.

To investigate on such criteria our starting point is the so-called "Giza diagonal" (Fig. 1). It is an ideal line which connects the south-east corners of the three Giza pyramids with good accuracy (a line between the G1 and G3 corners leaves the G2 corner about 20 meters to the north). The existence of this line tells us *two* important things. The first one is just that the site of Giza was inter-visible with Heliopolis and that this does not occur by chance: witness is the placement of the three monuments in accordance with such a Heliopolis-directed "diagonal". The importance of this observation has been enhanced by a comprehensive study of the inter-visibility between Heliopolis and the pyramids (Jeffreys 1998). In this study it is shown that the sites of the 4$^{th}$ dynasty pyramids whose owners first "declared an affinity" with the sun cult through their monuments and/or their names were chosen in such a way to be visible from Heliopolis. In the same study however it was also noticed for the first time that Abu Sir - another burial place of "solarised" kings - is *not* visible from Heliopolis: it actually stands just a few hundreds meters to the south of the last visible point. In spite of this, a Abu Sir "diagonal" - very similar to that of Giza - does exist. It connects the north-west corners of the pyramids built by Sahure, Neferirkare and Neferefre and it does indeed "point" to Heliopolis, in spite of the fact that the place is in itself invisible. The intentionality of this alignment – which was, therefore, realized using signals located half-way on the outcrop - has been controlled beyond any possible doubt using transit survey measures (Verner 2002).

The "invisibility" of the Abu Sir diagonal – which governed from the very beginning the topography of three pyramid building sites there - is a first hint that something of importance has yet to be understood in all this scenario. Another hint comes from a simple observation. In consequence of the alignment, looking from Heliopolis (or from any other point of the axis) the images of the Giza pyramids create a prospective effect: they "contract" on each other and merge into that of the Great Pyramid. Therefore, it can be said that the G2 and G3 pyramids were planned respecting *two* constraints (Magli 2009):

1) The chosen place is in plain view from Heliopolis
2) In this place, the two buildings G2 and G3 are located in the unique available positions which are actually *invisible* from the town, due to the presence of G1

While point (1) can be explained by the will of the ruler of writing in the stones his close connection with the Sun God, point (2) seems to await for an explanation. As far as Abu Sir is concerned, the situation is even more intriguing. Indeed, here we have

1) The chosen place is *the first available going south* to be invisible from Heliopolis
2) In this place, the second and the third successive buildings are disposed in such a way to be "symbolically" invisible from Heliopolis. In other words, looking along the "Abu Sir diagonal" they become almost invisible (in the second phase of construction S2 was re-planned to be higher than S1).

To explain this chain of occurrences, I propose that they correspond to a way of interpreting and modeling the sacred landscape which ruled the placements of the pyramids of the "solarised" kings up to the end of the 5$^{th}$ dynasty. This model, which I propose to call *symbolic invisibility*, required the pyramids of dynastically-related pharaohs to be built along a line of sight connecting a key element of the layout and then ideally extending up to the temple of Heliopolis. In this way, the direct lineage of the Pharaoh from the Sun God was "actualized" on the ground (see the Discussion section for further insights).

To illustrate how the model works, we will now re-run the succession of the kings mentioned earlier, putting in evidence if and how the location of their respective monuments fits into the framework.

1) Khufu selects Giza as the building place for his pyramid. The Giza plateau is certainly suitable for the enterprise, but there are hints that the place might have been chosen also because it was sacred to the sun since the previous dynasties.
2) The first to conform to the "symbolic invisibility" concept was Djedefre, Khufu's son. Indeed, if we plot the line which connects Heliopolis with his pyramid's south-west corner, we see that it crosses near the south-west corner of the Abu Roash pyramid A2, the so called Lepsius 1, which sits at the easternmost end of the Abu Roash hills. Rather than forming an Akhet sign (two coupled mountains with the sun setting at the center at the solstice) as has been recently suggested (Shaltout et al. 2007) these two pyramids thus form a "Abu Roash diagonal" pointing to Heliopolis.
3) After, Khafre decided to conform himself to the model as well, but returning to Giza; perhaps he appropriated and finished the construction of the second pyramid, but *in any case* he wanted his tomb to be attached "modestly" to a pre-existing project, that of Khufu, respecting the symbolic invisibility.

4) The very same choice was made by Menkaure, who arranged the layout of his funerary complex in order to harmonize with the pre-existing one. The dynastic lineage of Khufu was thus actualized at Giza by the succession – which actually *proceeds into the desert* - of the south-east corners of G1,G2 and G3
5) The un-attributed pyramid at Zawiet el Arian does not fit this scenario; some authors attribute it to the 3$^{th}$ dynasty (an attribution which of course would solve the problem here) but the similarity of its project with 4$^{th}$ dynasty pyramids is reported to be apparent[1] (for further discussion on this pyramid see sections 7 and 8).
6) Menkaure was succeeded by his son Shepsekaf, who broke the "solarised" tradition building his tomb as a Mastaba at Saqqara South, not far from Dashour and completely out of sight from Heliopolis (on this monument see also section 8).
7) After him, Userkaf actualized an ambitious building program, which consisted in exhibiting a return and a close link to old traditions of both the 3$^{th}$ and the 4$^{th}$ dynasties. Indeed, to show is closeness to the 3$^{th}$ dynasty kings he built his pyramid as close as possible to that of Djoser (perhaps inaugurating also a "topographical-dynastic diagonal" here, see Section 8). To exhibit explicitly his connection with Heliopolis and the "solarised kings" tradition of the 4$^{th}$ dynasty, Userkaf built the sun temple in Abu Gorab. Interestingly enough, the temple was constructed very near the southernmost available point of the west bank of the Nile from which Heliopolis *is* still visible (Kaiser 1956, Jeffreys 1998).
8) With Sahure we definitively return to "solarised" kings. According to the proposals of the present paper, we can infer that *the natural choice for Sahure's pyramid would have been Giza*, with the construction of a fourth monument aligned along the "Giza diagonal". However, building a pyramid complex even more far away in the desert with respect to Menkaure´s would have been nearly impossible. Thus, the architect had to find a new idea to allow his king to conform to the "symbolic invisibility" model. This idea was to place the pyramid of Sahure in the *first* available location in the south from which Heliopolis is not visible: Abu Sir.
9) Immediately after, Neferirkare inaugurated a new "dynastic diagonal" in this place, putting the north-west corner of his pyramid on the ideal line of sight to Heliopolis which crosses the corresponding corner of the Sahure pyramid (the apparently exaggerate distance between the monuments is due to the necessity of avoiding a slope located in between, see Krejci 2001). Although Abu Sir is invisible from Heliopolis, any person at the temple would have been aware, looking at the western horizon, that the brilliant obelisk of the Userkaf sun temple indicated the beginning of the sacred area, where the kings of the 5$^{th}$ dynasty decided to be buried, and that their pyramids were actually disposed along a "diagonal" pointing to Heliopolis although "modestly" in an invisible way.
10) The fact that also the unfinished monument of Neferefre is aligned on the same diagonal is a strong hint to the idea that this king was the first successor of Neferirkare (Verner 2001). After him, his successor Shepseskare had to confront to the same problem which faced Sahure at Giza, namely, it was impossible to "attach" his pyramid to the pre-existing 3-pyramids "diagonal" without going very far in the desert. Thus, the architect of the pharaoh probably planned the pyramid in the space left between Userkaf's sun temple and Sahure's pyramid, remaining in this way not just invisible, but "symbolically invisible" from Heliopolis.

## 5. The Niuserre project and the meaning of the Sun Temples

The problem of finding a *symbolically adequate* place for the pyramid of the next king, Niuserre, became dramatic. In order not to go too far off in the desert the planners of his monument found no other way to exhibit the lineage of the king than placing his pyramid to the east side of an existing one, that of his father Neferirkare, a quite unique example of "intrusive" design. In this way - although it may seem incredible - they also managed to inaugurate yet another "diagonal" (Fig. 2). This "secondary" Abu Sir diagonal is a line (first discovered by Lehner 1985b) which "dynastically" connects the south-east corner of Niuserre's pyramid with those of the pyramids of Neferefre and Neferirkare and touches also the corner of the Mastaba of Ptahshepses, a very important personage who became a son-in-law of Niuserre. However, the main problem was still standing namely, the fact that the connection of the king with the sun god was not "symbolically" expressed by his funerary complex. We can therefore speculate that perhaps exactly this was the reason which inspired the construction by Niuserre of his own Sun Temple, located north of the Userkaf's one and therefore in plain view of Heliopolis. In this way (and although it may really seem incredible) his architects also managed to inaugurate yet another "diagonal". This diagonal is - at least to the best of the author's knowledge - presented here for the first time. It is a 7 km line oriented ~45° south of east which originates from the center of the Great Pit in Zawiet el Arian and intersects the south-west corner of the 3$^{th}$ dynasty pyramid in the same site. It then crosses over Abu Sir intersecting with impressive precision the basis of the obelisk in the Niuserre Sun Temple, passes over the perimeter of Userkaf's temple, and ends at the apex of the Niuserre pyramid before "fading" into the area of Memphis.

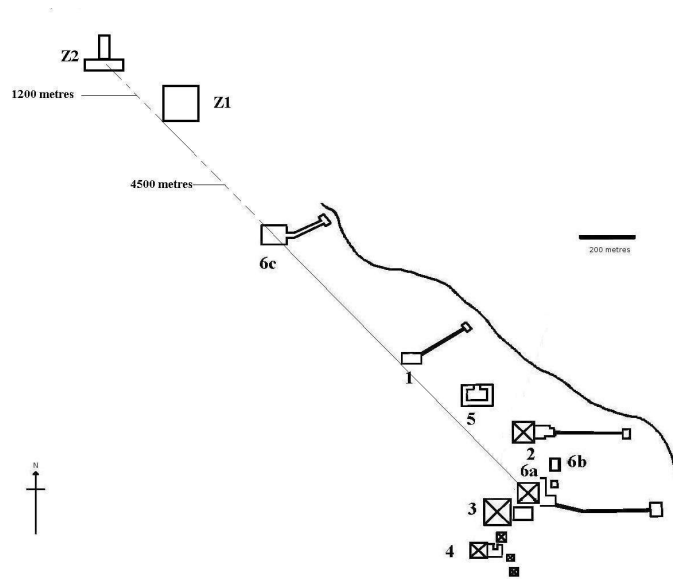

**Fig. 6 The Zawiet el Arian "diagonal"**

The function of this diagonal is thus to exhibit a close link between the funerary complex of the king and his Sun Temple. Actually, the true "meaning" and origin of the Sun Temples has never been explained satisfactorily. First of all, according to existing texts of the epoch, it seems that also Sahure, Neferefre, Neferirkare and Menkahour constructed one such monuments. However, *these temples have never been found.* As a consequence, some scholars have proposed that the texts actually refer to renewals made by such pharaohs to Userkaf's temple, and thus that the unique other sun temple effectively constructed after Userkaf's was that of Niuserre (for a discussion see Verner 2002, Nuzzolo 2007 and references therein). Actually, at least in the present author's view, it is almost impossible that the missing temples were completely dismantled as to safely escape to the archaeological investigation of the very restricted area in which they should have been located, since it is obvious that they had to be on the west bank and in view from Heliopolis (the idea, sometimes put forward in the past, that they could have been located on the east bank is – at least in the author's view – definitively incompatible with their symbolic meaning of "counterparts" of Heliopolis).

Clearly, the approach presented in this paper supports the thesis that the missing temples (or at least those of Sahure, Neferefre and Neferirkare, whose pyramids are known) never existed. In fact, I propose that Niuserre constructed a new temple near Userkaf's one *because* his pyramid was not symbolically linked to Heliopolis, being out of the Abu Sir diagonal. If this theory is correct, then at least one of the symbolic meanings of these temples can be further specified as follows: they were *witness of the royal link with the Sun God whenever this link was not explicitly stated by the location of the funerary complex of the king*. The theory proposed thus points to an extremely close connection between the construction of sun temples and that of the funerary complexes. Actually, the architectural complex composed by the "standard" elements Valley Temple/Causeway/Funerary Temple finds its first complete expression at Giza (in part it is already present at Meidum and Dashour). It starts therefore together with the "solarised kings" funerary complexes. If, as it seems likely (Quirke 2001) the Sun Temples were a sort of replica of the Heliopolitan one, then also the pyramid's temples complex - with their similar "two buildings + causeway" scheme - might have been conceived on the same basis. In this connection it may be noted that geological hints at a slight (a few centuries) pre-dating of parts of the Sphinx complex and Khafre mortuary temple to earlier dynasties exist (Reader 2001). If this - currently much debated, see e.g. Vandecruys (2006) - hypothesis will be confirmed by future research then these structures might refer to a earlier "sun temple" visually connected with Heliopolis, where the sun cult is known to be active at least from the 2$^{th}$ dynasty.

**6. The "solarised" kings and the pyramid's geometry**

As already mentioned, Niuserre was the last king to build his funerary complex in Abu Sir. The successive kings of the 5$^{th}$ dynasty, Menkahour, Djedkare and Unas, moved to Saqqara[2] as well as the very last kings who constructed pyramids in the Old Kingdom, those of the 6$^{th}$ dynasty (Section 8).

| | |
|---|---|
| KHUFU | 14/11 |
| DJEDEFRE | 14/11 |
| KHAFRE | 4/3 |
| MENKAURE | 5/4 |
| USERKAF | 4/3 |
| SAHURE | 6/5 |
| NEFERIRKARE | 4/3 |
| NEFERENRE | ? |
| NIUSERRE | 14/11 |
| MENKAHOUR | ? |
| DJEDKARE | 14/11 |
| UNAS | 3/2 |
| TETI | 4/3 |
| PEPI I | 4/3 |
| MERENRE | 4/3 |
| PEPI II | 4/3 |

Table 1. The slopes of the pyramids of the 4$^{th}$, 5$^{th}$ and 6$^{th}$ dynasties starting from Khufu's great Pyramid.[3]

The end of the time of the solar kings is perhaps reflected not only in the definitive abandon of the solar suffix and of any topographical connection with Heliopolis, but also in a "standardization" in the geometry of the pyramids themselves. To explore this issue, we refer to Table 1, in which the slopes of the pyramids from Khufu onward are reported in terms of the likely rational tangent which was used to construct them. With the exception of the very steep slope of the pyramid of Unas (a likely reason for such a slope will be discussed in Section 8), these values vary between 50° 12' (Sahure) and 53° 10' (Khafre). It has been recently noted that these slopes roughly correspond to the height of the sun when was passing the first vertical at the summer solstice during the Old Kingdom. This means that a spectacular illumination effect took place for a few days including the summer solstice: the west and south face were almost suddenly and fully illuminated by the ascending sun at the moment it crossed the first vertical (Belmonte and Zedda 2007). This observation furnishes an additional link between the pyramids and the solar cult and leads to speculate that different slopes were experimented in the context of this "sun and shadow" effect. Among these slopes, that of the second Giza pyramid is the unique which corresponds to a Pythagorean triangle (3-4-5, all integer legs). Whether this triangle was considered "sacred" - as passed on by much later, classical writers - or not, there is no doubt that Pythagorean triangles were the best suited for construction, because the builders could immediately check the correctness of the slope of the casing blocks. From the table it is apparent that, with the end of the "time of the solar kings" the slope of the pyramids becomes a fixed standard, actually the most suitable among those previously experimented, i.e. the 4/3 slope.

7**. Possible astronomical references of the Heliopolitan "diagonals"**

The "symbolic invisibility" framework provides a feasible explanation for the "diagonal" disposition of the successive plans of the monuments of the "solarised" kings. However, no doubt the funerary cult comprised, besides the "solar", also a "stellar" component. This is shown, for instance, by the accuracy of orientation to true north of the pyramid's bases and by the orientation of the shafts in the Khufu pyramid, as well as being confirmed slightly later by the Pyramid Texts (see Magli 2009, Magli and Belmonte 2009 for complete overviews on this subject). It is, therefore, worth analyzing if the choice of the sites may have had also a "nocturnal" connection with Heliopolis, where, as is well known, astronomy was practiced (clearly, if existing, such a connection might have been conceived and used also as an help in tracing the "diagonals"). To investigate this issue we check if the azimuths of the "diagonals" correspond to the setting of bright stars as viewed from Heliopolis (Fig. 5), and the results are as follows:

1) The "Abu Roash diagonal" (~28° south of west) is already known to be an astronomically significant sight line (see discussion in Shaltout et al. 2007). It points quite precisely to the winter solstice sunset; it also points to the setting of Sirius (within one degree if Djedefre accessed in 2528). Perhaps this was connected to the name of the pyramid, which was "Djedefre is a star *Sehedu*" (we do not know the meaning of the term *Sehedu*).
2) The "Giza diagonal" (~45° south of west) points to the setting of the brightest part of the Milky Way. At those times an observer looking from Heliopolis would have seen the stars of the Southern Cross, followed by the very bright stars of Centaurus, "flow" together with the great celestial river and disappear from view behind the apex of the Great Pyramid. In particular if Khufu is dated 2551 then the bright star Rigil sat in optimal alignment with the diagonal.
3) The "Abu Sir diagonal" (~71° south of west) corresponds to the setting of the star Canopus. As already mentioned, the site was "just" invisible from Heliopolis, and therefore the star was viewed to set over the outcrop of the Cairo citadel. In particular if Sahure is dated 2458 the agreement is within one degree.

Of course, such occurrences might be accidental. In support of their being perhaps not accidental the following argument can be cited. All the above mentioned stars were *Decans* (Belmonte 2001; about Canopus see however also Belmonte and Lull 2006). The Decans were 36 celestial objects whose heliacal rising happened in successive "weeks" of 10 days. The (ritual) use of the Decans to count the hours of the night is documented in Egypt from the 9th Dynasty (2154 BC) but probably dates before. We thus have three decanal stars or groups of stars - Sirius, Crux-Centaurus, and Canopus - which respectively sat in (approximate) alignment with the three "diagonals" - Abu Roash, Giza, Abusir - when viewed from Heliopolis. Further, these stars had heliacal settings (the last day in which the star is barely visible, immediately after sunset, and then sets) in successive periods going from north to south (Table 2). It may be noticed that the unfinished pyramid at Zawiet el Arian fits fairly well in the above described astronomical picture. Indeed, the sight line between Z2 and Heliopolis bears an azimuth ~56° south of west. This azimuth, besides identifying the rising of the bright star Arcturus over Heliopolis (Shaltout et al. 2007), points to the setting of the decanal star Fomalhaut as seen from the sacred city (within one degree at a fiducial date of, say, 2500). Further, the Heliacal setting of Fomalhaut in early November fits in between the dates for Crux-Centaurus (Giza) and Canopus (Abu Sir).

| LOCATION | RULER | DATE | AZIMUTH | STAR | HEL. SETTING (JUL.) | MAP |
|---|---|---|---|---|---|---|
| ABU ROASH | DJEDEFRE | 2528 BC | 28 S. OF WEST | SIRIUS | EARLY MAY | A2 |
| GIZA | KHUFU | 2551 BC | 45 S. OF WEST | RIGIL | END JULY | G1 |
| ZAWIET EL ARIAN | ? | 2500 BC | 54 S. OF WEST | FOMALHAUT | EARLY NOV. | Z2 |
| ABU SIR | SAHURE | 2458 BC | 71 S. OF WEST | CANOPUS | EARLY APRIL | S1 |

Table 2. The correspondence between the alignments of pyramids of the 4th and 5th dynasties and the decreasing setting azimuths of decanal stars as viewed from Heliopolis at the (likely) time of construction of the monuments

**8. The Unas' project and the Saqqara diagonal.**

The "symbolic invisibility" thus led, at least according to the thesis maintained here, to the birth of three "ideal lines" which connected Heliopolis with the pyramid fields of Abu Roash, Giza and Abu Sir respectively. With Niuserre, two further ideal lines were conceived, which however lost the "solar" character. Interestingly enough, the idea of "dynastic" alignments (representing lineage, or closeness of religious ideas, or more simply hints to past traditions) remained up to the end of the Pyramids age. It is, indeed, clearly visible also at Saqqara.

The existence of a "Saqqara diagonal" has been discovered by M. Lehner (1985b), who however did not attempt to discuss its meaning. It is a line oriented roughly SW-NE (it is difficult to ascertain its azimuth precisely, because the corners of the pyramids are not cleared, but it can be estimated as being ~48° north of east) might have been inaugurated by Userkaf. He placed his pyramid near the north-east corner of Djoser's precinct. In this way he aligned the south-east corner of his pyramid with the south-east corner of Djoser's pyramid and the north-west corner of Sekhemkhet's unfinished pyramid. It should be noted however that perhaps this "inhomogeneous" alignment was not intentional: the Sekhemkhet's pyramid originally had a high enclosure wall, and the pyramid itself might have been already buried at Unas' time (Fakhry 1974). Userkaf might have just chosen the best position - i.e. the one nearest to edge of the plateau - among all the possible locations near Djoser's precinct. Who, however, *certainly* did not choose a good position is king Unas, the last king of the 5th dynasty. He obliged his architects to arrange the diagonal of his pyramid on the line connecting the south-east corners of Djoser's and Userkaf's pyramids. There can be *no doubt* on the fact that this alignment was designed for symbolical, and not practical, reasons. To achieve the goal the architects were in fact obliged to build the pyramid near the south-west corner of the precinct of Djoser, thus very far in the desert. Consequently, they had to construct also a very long (more than 700 meters) causeway and, even worse, had to clear the zone near Djoser's south wall which was occupied by many pre-existing mastabas. Some of such tombs were thus interred, others even dismantled. Yet another problem was the height of the wall of the Step pyramid complex, which would have obstructed the view of the Unas pyramid to a person approaching the diagonal alignment; it is for this reason (at least in the author's view) that the pyramid was planned with a very steep slope, actually the steepest slope of all the Old Kingdom pyramids. To increase visibility the line was further traced along the diagonal of the pyramid, not along the south-east corner, again a unique case among the various "diagonals" which usually connect the same structural elements.

The placement of Unas' pyramid closely resembles that of Menkaure´s pyramid in Giza, to the point that the resemblance can hardly be considered casual. A similarity was already noticed some years ago by Goedicke (2001), who observed that an unobstructed line of sight connects the Userkaf pyramid with Khufu's. Actually, if similar connecting lines are traced between the summits of Djoser's and Unas' pyramids with the apexes of Khafre's and Menkaure´s monuments respectively, it becomes clear that the placement of Unas pyramid was conceived to realize a sort of (rough) copy of the arrangement of the Giza pyramid field. These lines are in fact about 14.5 Kms long (and therefore allowed a direct inter-visibility) and, although they are *not* parallel, their relative deviation stays within 2°.

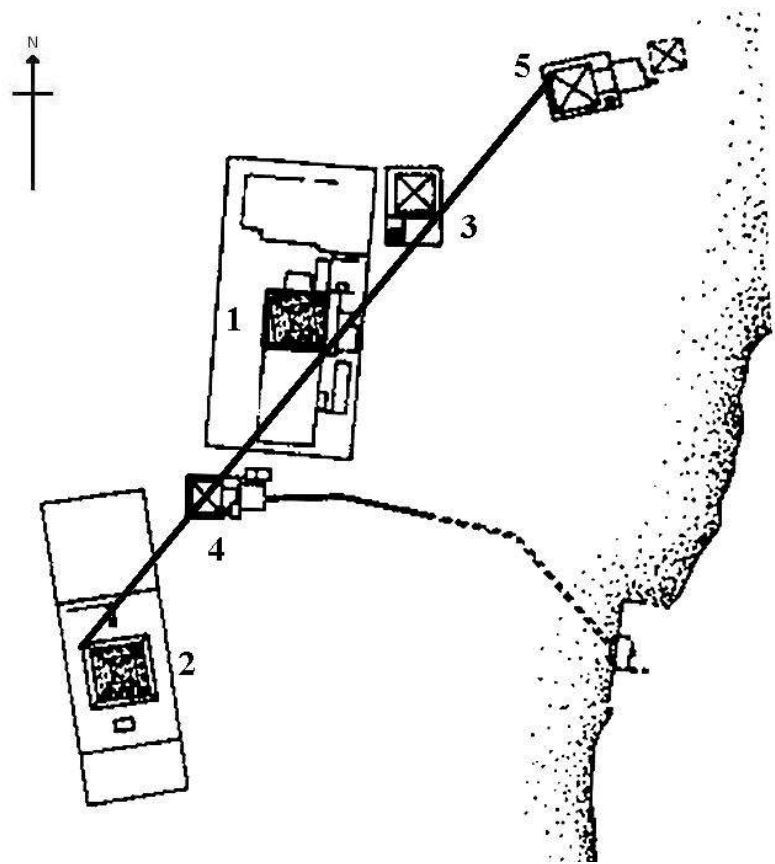

**Fig. 7 The "Saqqara diagonal" (numbering of the monuments in chronological order). 1- Djoser Step Pyramid, 2- Unfinished pyramid of Sekhemkhet, 3-Pyramid of Userkaf, 4- Pyramid of Unas 5-Pyramid of Teti**

Some years later, the Saqqara diagonal was further continued by Unas' successor Teti. A old-fashioned "symbolic invisibility" would have required Teti to build his pyramid in a position very far in the desert, along the diagonal to the south-west of the Unas one. This zone was partly occupied by the Sekhemkhet's unfinished complex, and/or the original symbolic meaning of the "diagonals" was lost; in any case Teti avoided this inconvenient position and actually choose to built his complex in the most favorable position along the Saqqara diagonal, aligning the north-west corner of his pyramid and placing it to the north-east of the Userkaf's one, very near the ridge of the Saqqara plateau. Prolongation of the line further north-east shows that it passes very close to the southern limit of the area of the early-dynastic mastabas (more precisely, the line passes some 70 meters south of the southernmost Mastaba 3507) and then crosses over the (un-excavated) zone of archaic Memphis. It would thus be interesting to ascertain if an ancient path coming from the valley was ascending on the Saqqara plateau from this direction, thus creating a visual "superposition" effect similar to that visible approaching Giza from Heliopolis. The likely existence of this path might also be of help in understanding another "topographical mystery", that of the location of Shepsekhaf's tomb. In fact this beautiful monument (which was probably meant to resemble the so-called Buto shrine, a sort of "bench" with straight ends) was located in the area between the Central Saqqara field and Dashour which was, at that time, free of significant buildings. However approaching the Saqqara Plateau in the above mentioned direction the view was unobstructed (since Userkaf and Teti pyramids were still to be built) and it was possible to see the king's tomb forming a sort of regular baseline for a double-mountains symbol "created" at the horizon by the two giant pyramids of Snefru.

The pyramid building site of Teti's immediate successor, Userkare, has never been individuated. After Userkare, the king Pepi I built his pyramid in a position in south Saqqara, very near the ridge of the plateau, while his successor Merenre moved further south-west (Fig. 8). In this way, he aligned the diagonal of his monument to that of Pepi I (also the position of the center of this pyramid was probably intentional, see next section). Unfortunately, however, we shall never know if Pepi II, successor of Merenre, would have liked to add his monument to this (supposed) new diagonal, because the corresponding position to the south-west would have been located in a Wadi (dried river), absolutely not suitable for building a pyramid. Perhaps as a consequence of this, he choose a position immediately south of the same Wadi, near Shepsekhaf's monument.

**9. Meridian alignments at Saqqara south and Dashour**

The french Egyptologist George Goyon proposed several "meridian" (i.e. north-south) alignments between pyramids and previously constructed monuments "belonging to an ancestor or sacred sites" (Goyon 1977); the most famous of Goyon's alignment is that of Giza with the sacred site of Khem (Letopolis) located due north. A search for meridian alignments in the pyramid fields effectively gives quite surprising results, which can hardly be attributed to a chance. Further, the idea of meridian alignments appears before Giza, because it is already present at Dashour. Indeed we have (in chronological order):

1. The Red and the Bent pyramid of Snefru are constructed in such a way that their west sides align with the two south corners of the so-called *Gisr el-Mudir*, a huge rectangular enclosure located southwest of the Step Pyramid and probably constructed by a king of the first two dynasties, perhaps Khasekhemwy (Goyon already noticed at least one of such alignments, which he refers to the "De Morgan" - the discoverer of the Gisr el-Mudir – precinct). This alignment actually helps to explain why the Red Pyramid was constructed so far in the desert, while the relative distance between the two remains something of a mystery.
2. The center of the valley temple of the Bent pyramid aligns with the apex of the Step Pyramid.
3. A meridian alignment connects the apex of Unas with the central axis of Shepsekhaf's Mastaba. On this line also the pyramid of Merenre was later constructed.
4. A meridian alignment connects the apex of Pepi I with that of Userkaf.

It appears, therefore, a possible intentionality in the successive planning of the Pepi I-Merenre pyramids, which ideally connects them with the "Unas' project". At a purely speculative level, one might think that it was Userkare to initiate this new pyramid field. Interestingly, recent excavations in the *Tabbet al-Guesh* area, namely the zone which separates the Saqqara central group from the Pepi I complex, gave many hints at the possibility that a pyramid has still to be discovered there (Dobrev 2008). This area is actually crossed by yet another possibly non-casual connecting line, since the northwest-southeast diagonal of Pepi I aligns with the south-east corner of the Gisr el-Mudir.

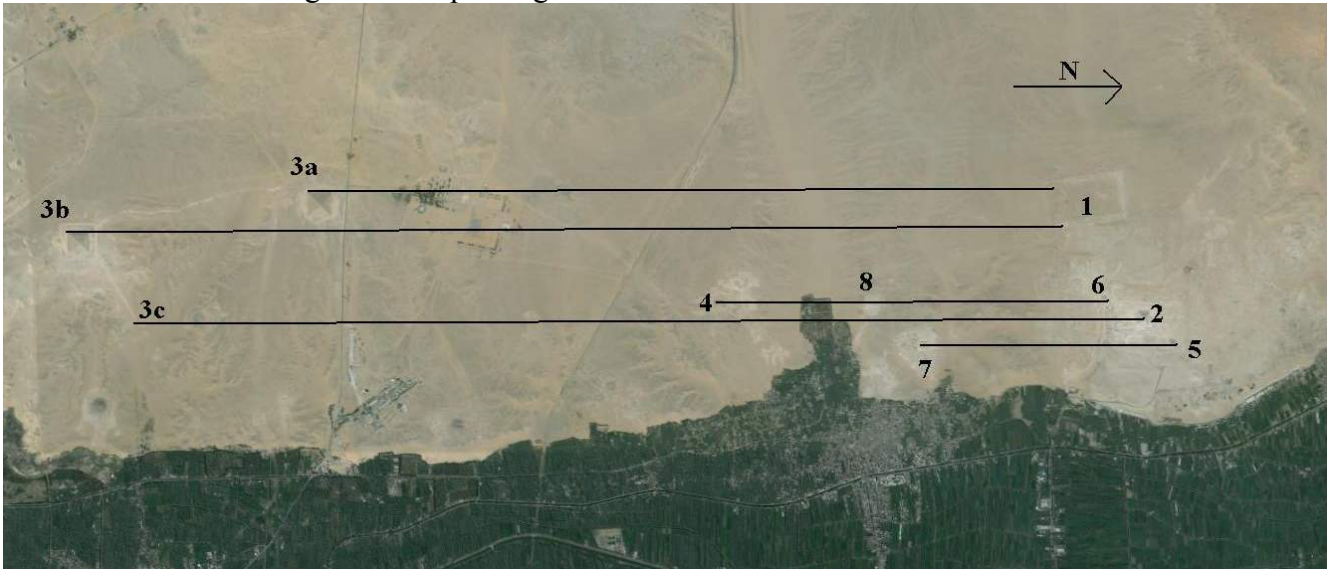

**Fig. 8 Meridian alignments between Dashour and Saqqara (numbering of the monuments in chronological order). 1- Gisr el-Mudir 2- Djoser Step Pyramid, 3a/3b/3c - Red Pyramid, Bent Pyramid, Valley Temple of the Bent Pyramid 4- Mastaba El Faraun 5-Pyramid of Userkaf, 6- Pyramid of Unas 7-Pyramid of Pepi I 8-Pyramid of Merenre**

## 10. Discussion

Mircea Eliade once said (1978) that the symbolism contained in the sacred space is "so old and so familiar" that it may become impossible to recognize it. The present work represents an attempt at individuating the existence of a general pattern which inspired the human-made landscape during one of the greatest seasons of human civilization. We can call this period the time of the solar kings, those great rulers who were "close to the Sun God" as the name of Sahure has it.

This period was inaugurated by Khufu. According to authoritative scholars indeed (Stadelmann 1991, Hawass 1993) Khufu depicted himself as the incarnation of Ra. In this connection Hawass notices that no relief is known showing Khufu making offerings to Gods, and perhaps this is not due to the lack of documentation but rather is in accordance with the new "cult" which would have equated the king with the god himself. If this is true, then the "solarised" kings would have had the necessity of showing in explicit, monumental terms their direct lineage from the Sun God in the funerary complexes. In the present paper, it has been proposed that the way in which this will was actualized was the intentional alignment of structural features of subsequent pyramids. This had the consequence of creating visual, prospective effects, in which the monuments progressively "merge" each other.

Of course, we cannot attempt here at developing a complete analysis of the possible consequences in terms of dynastic (or simply "inspiration") relationships between successive kings who decided to connect visually their monuments to those of previous pharaohs. However, at least a few observations in support of the non-randomness of the "topographical order" we are discussing here can be made. First of all, the ubiquitous (Giza, Zawiet el Arian-Abu Sir, Saqqara) existence of survey lines oriented quarter-cardinally in the pyramid fields is reflected in a fact which emerged very clearly from the recently completed survey of the orientations of virtually all the ancient Egyptian temples (Belmonte et al. 2009). The analysis of the data indeed led the authors to identify a series of "families" of orientation; among others (like e.g., of course, a "solstitial" family) also a "quarter-cardinal" family of orientations is clearly discernible, and comprises temples of any epoch.

As is well known, no contemporary textual evidence is available regarding the pyramid's projects and design. However, an echo of a Heliopolitan symbolism of "progressive ascendance" can perhaps be seen in a passage of the pyramid texts, where it is said (PT 307) "*My father is an Onite, and I myself am an Onite, born in On [=Heliopolis] when Ra was ruler*". Perhaps another echo appears in the traditional tales passed on in the Westcar papyrus (Lichtheim 1973). In this papyrus (dating to the Middle Kingdom) five stories which are told to Khufu by his sons are reported. In one such stories, the beginning of the 5$^{th}$ dynasty is directly linked to the sun in a quite intriguing way. Indeed, a magician called Dedi prophesies to Khufu that three sons of the wife of the chief priest of Ra in Heliopolis (and of Ra himself, at least according to some translations) will reign, the first of them being Userkaf. Khufu gets upset of this, but Dedi reassures the king that this will happen only after that Khufu's son and grandson will reign. Of course, Westcar has not to be considered as an historical document; however, the tale clearly alludes to a discontinuity between the 4$^{th}$ and the 5$^{th}$ dynasties by "re-scaling" the lineage of the kings directly to Heliopolis, exactly - one would be tempted to say - as it happens on the ground.

## Acknowledgments

The author gratefully acknowledges Juan Belmonte for a careful reading of the first draft of the present paper and many constructive comments.

[1] The monument lies inside a military base and is the unique site mentioned here that the author has not personally surveyed.

[2] The Menkhaour pyramid is identified, without certainty however, with the pyramid Lepsius 29 lying east of the Teti complex; the pyramid of Djedkare was built in a prominent position on the ridge in South Saqqara; the Unas pyramid lies in Central Saqqara and will be discussed at length in section 8.

[3] There can be little doubt about the fact that the slopes of the pyramids were chosen in the planning phase and controlled during the construction by means of rational tangents, i.e. choosing two integer numbers. We thus report here the "simplest" rational fractions (lowest possible numbers) giving a reasonable rational approximation (within 1%) of the measured slopes. It should be noted, however, that in *much later* sources (such as the Rhind Papyrus) a quite awkward way of measuring the tangent of an angle, called *seked,* appears. The seked is how much one has to "slide" the base in order to achieve an height of one cubit. This measure has the obvious drawback of being dependent on the unit of measure, while angles are of course non-dimensional - and therefore easy-to-handle - physical quantities. Although no proof of the use of the seked in the Old Kingdom exists, it is customary in the literature to try to express the slopes of the pyramids in this way (see e.g. Rossi 2003). Since the cubit was divided into 7 palms and the palm was divided into 4 fingers, even some eagerly simple slopes (such as 3/2) become utterly complicated when expressed into the *seked* system. There is no reason whatsoever to believe that the builders of the pyramids used this way of working, if only it is observed that workmen in charge of cutting casing block could use *arbitrary* units to check their work if the slope was instead given as a two-integers numbers information.